\documentclass[sn-mathphys]{sn-jnl}

\jyear{2021}%

\theoremstyle{thmstyleone}%
\theoremstyle{thmstyletwo}%

\theoremstyle{thmstylethree}%

\begin{document}

\title[Experimental investigation of the effect of ionization on the $^{51}$V(p,n)$^{51}$Cr reaction]{Experimental investigation of the effect of ionization on the $^{51}$V(p,n)$^{51}$Cr reaction}


\author*[1,2]{\fnm{Vincent} \sur{Lelasseux}}\email{vincent.lelasseux@eli-np.ro}

\author[2]{\fnm{Pär-Anders} \sur{Söderström}}
\author[2]{\fnm{Soiciro} \sur{Aogaki}}
\author[1,3]{\fnm{Konstantin} \sur{Burdonov}}
\author[2]{\fnm{Sophia} \sur{Chen}}
\author[4]{\fnm{Aleksandra} \sur{Cvetinović}}
\author[1]{\fnm{Emmanuel} \sur{Dechefdebien}}
\author[1]{\fnm{Sandra} \sur{Dorard}}
\author[1,5]{\fnm{Alice} \sur{Fazzini}}
\author[6]{\fnm{Laurent} \sur{Gremillet}}
\author[2]{\fnm{Marius} \sur{Gugiu}}
\author[1,2]{\fnm{Vojtech} \sur{Horny}}
\author[4]{\fnm{Matej} \sur{Lipoglavsek}}
\author[7]{\fnm{Sergey} \sur{Pikuz}}
\author[2]{\fnm{Florin} \sur{Rotaru}}
\author[1]{\fnm{Yao} \sur{Weipeng}}
\author[2]{\fnm{Florin} \sur{Negoita}}
\author*[1]{\fnm{Julien} \sur{Fuchs}}\email{julien.fuchs@polytechnique.edu}

\affil*[1]{\orgdiv{LULI-CNRS}, \orgname{CEA, UPMC Univ Paris 06: Sorbonne Université, École Polytechnique}, \orgaddress{\street{Institut Polytechnique de Paris}, \city{Palaiseau}, \postcode{91128}, \country{France}}}
\affil[2]{\orgname{IFIN-HH}, \orgaddress{\street{Str Reactorului}, \city{Magurele}, \postcode{077125}, \country{Romania}}}

\affil[3]{\orgdiv{IAP}, \orgname{Russian Academy of sciences}, \orgaddress{\city{Nizhny Novgorod}, \postcode{603950}, \country{Russia}}}

\affil[4]{\orgname{Jozef Stefan Institute}, \orgaddress{\street{Jamova cesta}, \city{Ljubljana}, \postcode{1000}, \country{Slovenia}}}

\affil[5]{\orgname{Marvel Fusion GmbH}, \orgaddress{\city{Munich}, \postcode{80339}, \country{Germany}}}

\affil[6]{\orgdiv{DAM, DIF}, \orgname{CEA}, \orgaddress{\city{Arpajon}, \postcode{91297}, \country{France}}}

\affil[7]{\orgdiv{Joint Institute for High Temperature RAS}, \orgname{Russian Academy of sciences}, \orgaddress{\street{Izhorskaya St}, \city{Moscow}, \postcode{125412}, \country{Russia}}}


\abstract{The investigation of the effects of average atomic ionization on nuclear reactions is of prime importance for nuclear astrophysics. No direct experimental measurement using a plasma target has been done yet. In this regard, we measured for the first time the neutron production of a (p,n) reaction in different states of ionization. The studied nuclear reaction was $^{51}$V(p,n)$^{51}$Cr. We measured a significantly lower neutron production than expected when the target was ionized, even when taking into account existing electron screening theory or the effect of the stopping power in the target on the injected proton beam. This experiment is a first step in the process to characterize the influence of ionization at astrophysically relevant energies.}




\maketitle

\section{Introduction}\label{sec1}

The effect of the plasma-electron screening on nucleosynthesis is a long-standing problem. Experimental data obtained with solid targets at astrophysically relevant energies, i.e. energies in the Gamow peak, is far from the theoritical predictions\cite{salpeter1954electron,strieder2001electron,spitaleri2016electron,lipoglavsek2020electron}. The latest experiments report screening measurements in solid targets an order of magnitude above the theoritical predictions\cite{cvetinovic2020electron}.

When it comes to plasma targets, there is simply no experimental data available to constrain the models\cite{casey2023towards}. Reliable theoritical predictions are difficult to produce since plasma behaviour is often very subtle and depends on a lot of parameters \cite{cereceda2000dielectric,hayes2020plasma}. However, such predictions are of prime importance since, for example, this screening effect should imply a change of a few percents in the nucleosynthesis reaction rates in the Sun but could reach six orders of magnitude for others astrophysical sites\cite{national2003frontiers}.

With the development of multi-PW lasers, such as Apollon\cite{papadopoulos2019first,burdonov2021characterization} or ELI-NP\cite{radier202210,tanaka2020current}, the particle beams which can be produced by condensing light on a target reach new milestones. In particular, the proton and neutron beams those facilities will be able to allow experimental nuclear physics to study problems which cannot be addressed by conventional facilities\cite{zamfir2014nuclear}.

In the work presented here, we took advantage of the possibility that high-power lasers can both produce a bright proton beam, and ionize targets at the same time. The protons being produced in bunches of picosecond timescale duration, the state of plasma target can be considered as a steady state. In this way, we could test the effects of ionization on a nuclear reaction, which we chose here to be a (p,n) reaction in vanadium.

To reach this goal, we developed a high efficiency neutron counter adapted to the high intensity laser environment \cite{lelasseux2021design}. As almost no neutrons are produced during the production of a laser-driven proton beam, (p,n) reactions are good candidates to measure cross sections in such harsh environments. As shown by Lelasseux et al. \cite{lelasseux2021design}, neutron measurement was partly delayed to avoid the electromagnetic background produced by the prompt laser-matter interaction.

The paper is organized as follows. In Section 2 , we detail the experimental setup and the parameters of the shots that were performed. In Section 3, we detail the analysis of the proton and neutron signals that were recorded in order to retrieve the actual number of neutrons emitted each shot. Section 4 presents the result of this analysis, mainly the unexpectedly low number of neutrons measured when the vanadium target was in a plasma state. Section 5 discusses possible reasons behind our observation and Section 6 concludes this study.

\section{Experimental setup}\label{sec2}
\subsection{Setup of the lasers and targets used in the experiment}

The experiment was performed at the LULI2000 facility located at École Polytechnique in Palaiseau, France. It took place in the experimental area \#1 using the MILKA experimental chamber. During this experiment, we used two laser beams. The first one, called the North beam, could deliver an energy up to 1 kJ at 1.053 $\mu$m wavelength in the ns duration regime. The second one, called the South beam, could deliver up to 60 J in 1 ps duration also at 1.053 $\mu$m wavelength.

\begin{figure}[ht!]
       \centering    
       \includegraphics[width=0.7\textwidth]{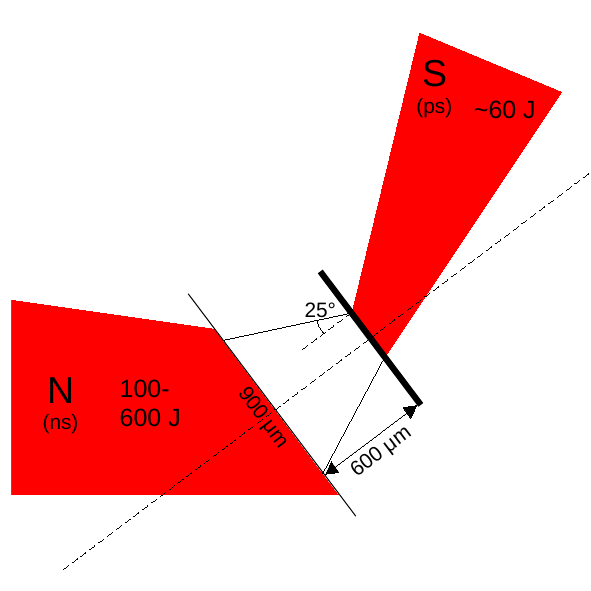}
       \caption{Sketch of the targets setup with angles and distances between them. The cone between the two targets represents the generated proton beam that also extends behind the vanadium target.}
       \label{fig:target_setup}
\end{figure}

We used a two targets setup shown in Fig.\ref{fig:target_setup}. The primary target was a 23 $\mu$m thick PolyEthylene Teraphtalate (PET) sheet with a 300 nm aluminum layer on the front face to enhance the laser absorption. It was used to produce a proton beam driven by the South laser beam through the Target Normal Sheath Acceleration (TNSA) \cite{macchi2013ion} mechanism. By changing the size of the focal spot on the target, thus changing the laser intensity, it was possible to modulate the maximum energy of the spectrum obtained from the TNSA, without modifying the spectrum otherwise\cite{foccutoff}.

The proton beam impinged onto a secondary target in which the (p,n) reactions occurred. In order to try to investigate the reactions at proton energies astrophysically relevant, we decided to use vanadium, due to the $^{50}$V stable isotope having a 255.78 keV positive Q-value and vanadium having the highest (p,n) cross-section \cite{endf} among this kind of materials. We decided to use a thin vanadium target. Indeed, a thin target ensures that the proton beam can be analyzed by a spectrometer placed downstream due to the scattering and stopping power being negligible in such a case. A thin target will also ensure that the heating, following the ns-duration laser irradiation, is as homogeneous as possible. Neutron production calculations and MULTI, a 1D radiation-hydrodynamic code \cite{ramis1988multi}, simulations were used to determine that a 1 $\mu$m  thickness was a good compromise between having the maximal mean ionization in the target, which required a thin target, and generate enough neutrons for us to be able to measure them, which would favor using thick targets.

The distance between the two targets is also the result of a compromise between having the targets close enough to limit the angular dispersion of the diverging proton beam \cite{bolton2014instrumentation}, and far enough for the x-ray emission of the vanadium target during its heating by the North beam, not to blow the primary target and prevent the proton beam generation \cite{mackinnon2001effect,fuchs2007laser}. We decided to set the minimum distance to fulfill the second condition at 600 $\mu$m. Hence, considering a maximum proton beam full divergence of 50°\cite{bolton2014instrumentation}, this gives us a disk of the projected proton beam with a 560 + X $\mu$m diameter on the secondary target, X being the diameter of the focal spot on the primary target. In practice, for the shots discussed below, we set X = 150 $\mu$m. To make sure that the ionized section of the target encompasses the proton beam, the North beam focal spot diameter was set to 900 $\mu$m.

\subsection{Diagnostics setup}

In order to maximize the angular coverage of the neutron detector, we limited the overall number of diagnostics to three. First, to monitor the ionization degree of the target, we used a FSSR\cite{FSSR} x-ray emission spectrometer. Second, to monitor the proton energy spectrum, we used a magnetic spectrometer\cite{cowan1999high}.

\begin{figure}[ht!]
       \centering    
       \includegraphics[width=0.7\textwidth]{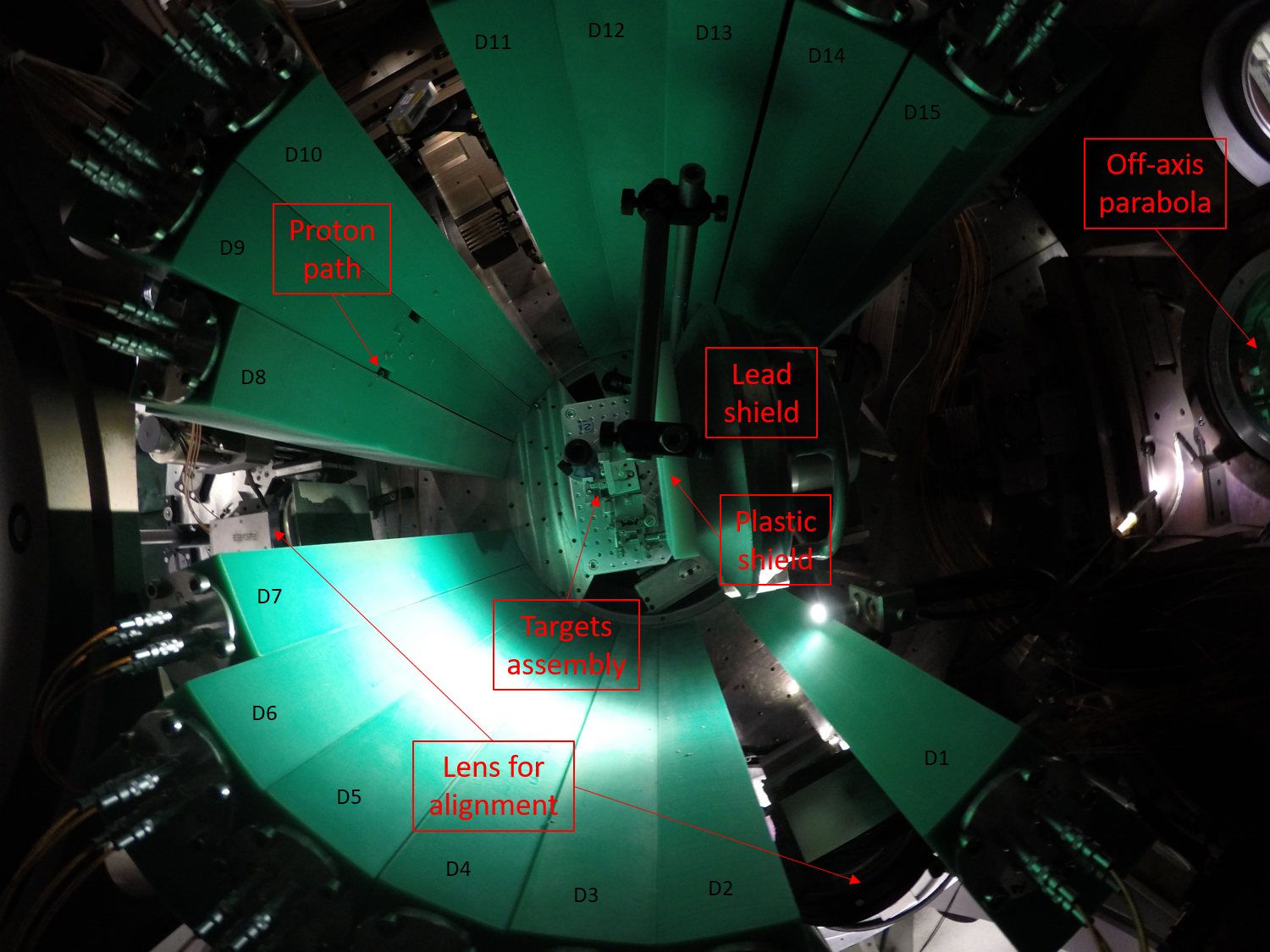}
       \caption{Top-view of the setup in the experimental chamber. The numbers of the detector modules are written in black and some parts of the setup are in red.}
       \label{fig:diag_setup_top}
\end{figure}

\begin{figure}[ht!]
       \centering    
       \includegraphics[width=0.7\textwidth]{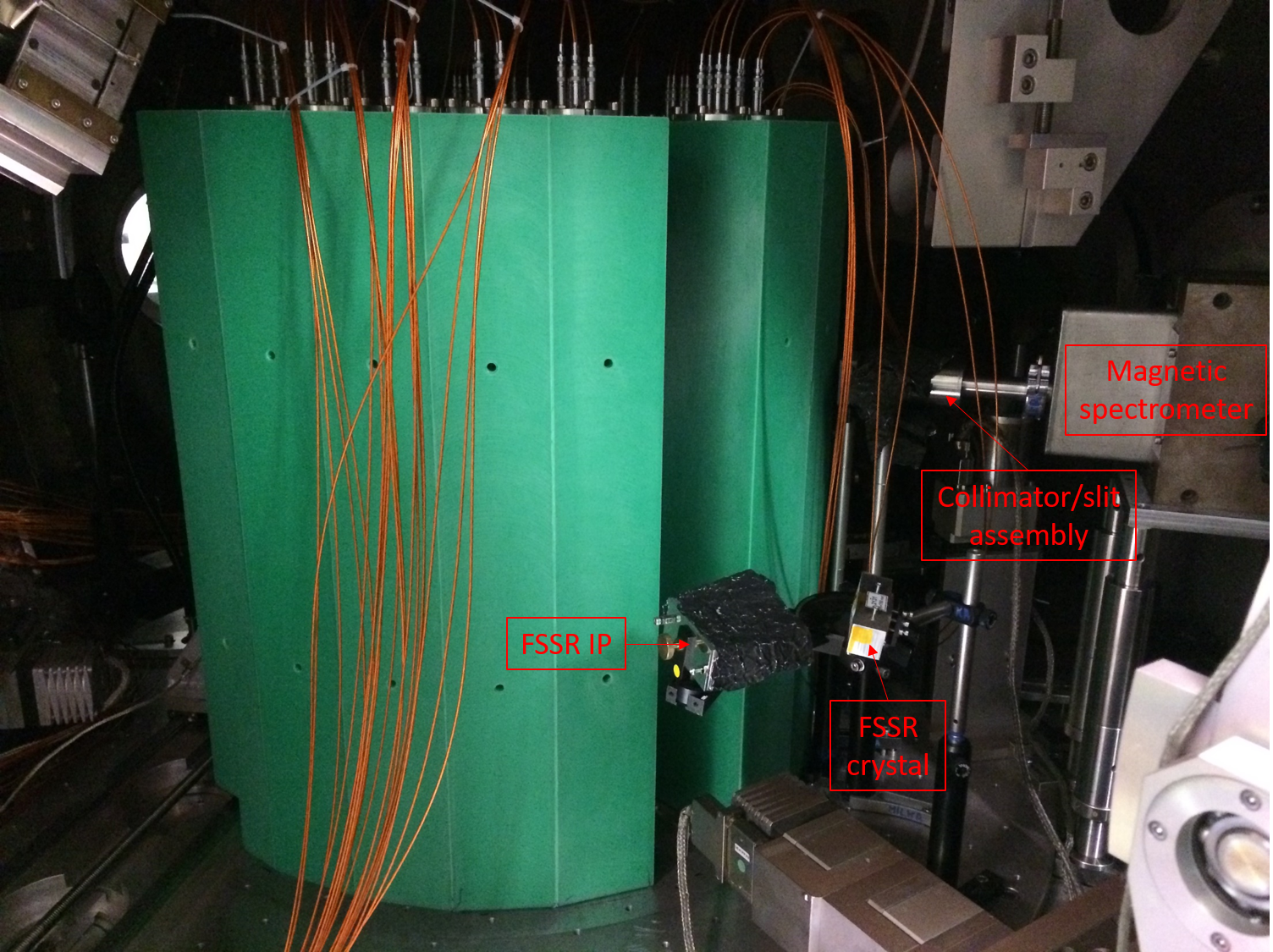}
       \caption{Side-view of the setup in the experimental chamber. Some parts of the setup are labelled in red. FSSR stands for Focusing Spectrometer with Spatial Resolution and IP for Imaging Plate.}
       \label{fig:diag_setup_side}
\end{figure}

Third, to measure the neutron emission, we used a scintillator-based neutron counter \cite{lelasseux2021design} whose dimensions have been determined taking into account constraints due to the experimental chamber size, the laser paths and other elements in the chamber, and using Geant4 simulations\cite{Geant4}. A top-view and a side-view of the placement of those diagnostics are respectively presented Figs.\ref{fig:diag_setup_top} and \ref{fig:diag_setup_side}. The biggest hole in angular coverage in the right side of the top view is due to the path of the picosecond laser beam, and especially to accomodate a lead shield which has the purpose to protect the experimental and technical teams from the proton beam which comes out from the front face of the PET target\cite{mckenna2004characterization}. To prevent those protons to reach the lead shield and produce a massive amount of unwanted neutrons, a plastic shield has been installed between the target setup and the lead shield. Another angular opening on the top left side of the top view (see Fig.\ref{fig:diag_setup_top}) has been created for the nanosecond laser beam path. For alignment purposes, it was also necessary to leave two spaces on the opposite side of the laser paths, here on the left side and on the bottom right side of the top view. Finally, a hole was machined in one of the neutron detector modules, to allow a part of the proton beam to go through it and reach the magnetic spectrometer placed further away. All the neutron detector modules had a 60 cm height except for units 7 and 8 for which the height is 45 cm. The target chamber center at which the targets were shot was at a height of 40 cm from the chamber breadboard.

\subsection{Shot sequences}

Four types, or series, of shots were performed during the experiment. First, in order to characterize the proton beam that could be generated and observe the neutron detector response, we performed shots without the secondary vanadium target. These are called "background" shots in the following. Second, we added a solid vanadium target on the path of the protons, as shown in Fig.\ref{fig:target_setup}. These are called "cold" shots in the following. Third, the vanadium target was ionized by the North beam. These are called "hot" shots in the following. Fourth and last, to check the influence on the detector of the North beam used to heat the secondary target, we also performed a number of shots only with the North beam.

\section{Analysis}\label{sec3}

Typical traces recorded by the neutron counter are shown in Figs.\ref{fig:trace_examples}. In these traces, we can observe that the unresponsive time in the PhotoMultiplier Tubes (PMTs) that are part of the neutron counter was much higher than what we could expect from the tests we run during other high intensity experiments\cite{lelasseux2021design}. This unresponsive time depending strongly on the PMT and its position, the dead time ranged from 100 $\mu$s (see detector 15\_D in Fig.\ref{fig:trace_examples}d) to more than 1 ms in the worst cases (see detector 7\_U in Fig.\ref{fig:trace_examples}c), 1 ms being the recording duration of the signal. To cope with this unexpectedly long unresponsiveness and the strong background, we established the analysis procedure detailed below, in order to ensure the nature and validity of our measurements.

\subsection{Proton analysis}

First of all, to be able to compare equivalent shots, we quantitatively analyzed the emitted proton spectrum for each shot with a procedure presented in Ref.\cite{lelasseux2022detection}. An example of the spectra are shown in Fig.\ref{fig:hot_d15}a. All the proton spectra present an exponentially decreasing behaviour, characteristic of TNSA, with a maximum energy cutoff around 16 MeV.


\subsection{Neutron analysis}
\subsubsection{PMT response to gamma flash/EMP}
\begin{figure}[ht!]
      \centering
       \begin{subfigure}[b]{0.49\textwidth}
            \centering
            \includegraphics[width=\textwidth]{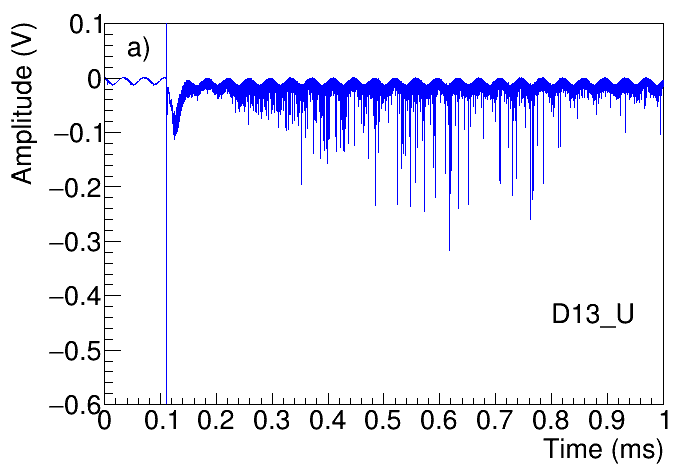}
       \end{subfigure}
       \begin{subfigure}[b]{0.49\textwidth}
            \centering
            \includegraphics[width=\textwidth]{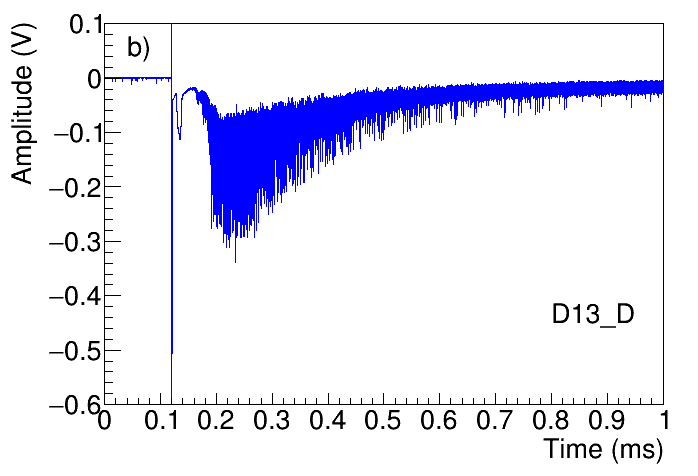}
       \end{subfigure}
       \begin{subfigure}[b]{0.49\textwidth}
            \centering
            \includegraphics[width=\textwidth]{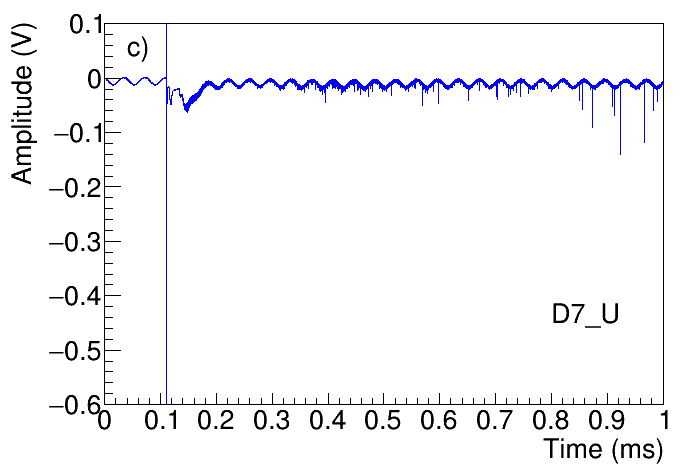}
       \end{subfigure}
       \begin{subfigure}[b]{0.49\textwidth}
            \centering
            \includegraphics[width=\textwidth]{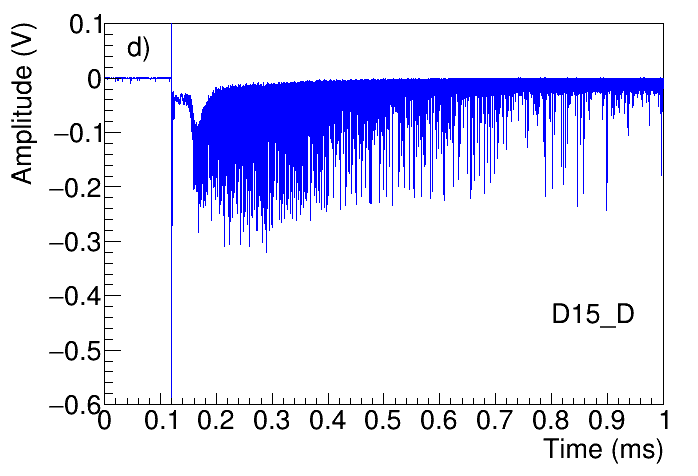}
       \end{subfigure}
       \caption{Examples of traces obtained from different detector units and PMTs during shot 28. During this shot, the proton energy cutoff was 16.31 MeV, a 1 $\mu$m solid vanadium target was in place and the North beam was not shot. "U" corresponds to PMTs on the up side of the modules and "D" to the ones on the down side.}
       \label{fig:trace_examples}
\end{figure}

We are then able to compare the neutron signals for shots having equivalent proton beams. To do so, we have to analyze properly the traces recorded by the pair of PMTs, working in coincidence, from each detector module. However, as mentioned above, the unresponsive time from most of the PMTs was longer than expected, as can be seen in Figs.\ref{fig:trace_examples}. The PMTs which are the closest to the picosecond laser direction, i.e. of detector units 6, 7 and 9 are more affected and cannot be used for quantitative analysis. The detector unit 8 had an issue during its construction and was not functioning. Only the PMTs of detector units 1, 14 and 15 have traces that we were able to analyze for all shots of interest. Taking the position of the lead shield into account, we suspect that the energy deposited in the scintillators by the gamma flash, induced by the prompt South beam/PET target interaction, is mainly responsible for the saturation observed after the first main peak around t = 0.12 ms.

From a closer look at the behaviour of those traces (see D13\_U in Fig.\ref{fig:trace_examples}a), we see that, after the unresponsive time, the maximum peak height increases to values between 250 mV and 300 mV. We note that this behaviour is abnormal since the mean peak height for a neutron detection is around 50 mV in normal conditions\cite{lelasseux2021design}. Following this, the maximum peak height decreases on the scale of a few hundred $\mu$s, as shown with the trace of the down PMT from detector unit 13 in Fig.\ref{fig:trace_examples}b. For some PMTs, as shown in Fig.\ref{fig:trace_examples}c for D7\_U, the saturation is so important that the 1 ms long recording time is shorter than the recovery time. For other PMTs, as D15\_D, the peak height remains at the higher than expected value for the whole recording time, as shown in Fig.\ref{fig:trace_examples}d.

As it can be deduced from Figs.\ref{fig:trace_examples} and the complete set of traces, the duration of the time window during which the maximum peak height is first at the noise level, and then increases to reach a maximum value, seems to mainly depend on the position of the PMT. The light distribution between the two coupled PMTs depends strongly on which scintillator cylinder the energy deposition takes place in, as detailed in Ref.\cite{lelasseux2021design}. Hence, PMTs on the upper side should indeed receive more light than the ones on the down side.

However, the time it takes for this maximum peak height to decrease seems to be PMT dependent. We can draw two conclusions from this. First, since for some PMTs, the maximum peak height eventually comes back to what we can expect from a regular neutron capture event, most of those peaks probably correspond to neutron capture events, even though their height is higher than expected. Second, the characteristics of the neutron peak in the energy histograms will probably be different than what is expected, on the base of the calibration using a source \cite{lelasseux2021design}. Indeed, the peaks which have a higher height than expected will most probably also have higher integrals and, hence, higher associated energies. Then, the best case scenario is that the mean peak height stays constant and high for both coupled PMTs. This way, the events corresponding to a neutron in the energy deposition histogram, i.e. the neutron peak, should be located at a higher energy than during the calibration. For all other cases, the maximum peak height varies in time for at least one of the two PMTs and hence the definition itself of the neutron peak is more complicated. In conclusion, only the units for which the mean peak height stays high for both PMTs, will produce energy deposition histograms with a clear neutron peak. These are the units that we will use for the following analysis.

For the shots we conducted using only the North beam to ionize the vanadium target, we observe that the North beam generates only electronic background. No angular or U/D dependence can be seen when we examine all the different traces. We do not observe any 200 mV-high peak during the shots without the South beam. Hence, the addition of the North beam should not have a significant effect, apart from a possible higher electronic noise depending on the PMT.

\subsubsection{Trace analysis}
First, we need to perform a temporal selection since the PMTs are unresponsive or not responsive in a satisfying way for some time after the shot. According to Geant4 simulations done using the actual experimental setup and detailed later, we know that the neutron detection rate should decrease according to the sum of two exponentials. In the range of neutron energy we are interested in, shown in Fig.\ref{fig:neutron_spectrum_shots}a, the two moderation components have exponential time constants of around 12 and 110 $\mu$s. Taking the dead time duration into account, which lasts at least 100 $\mu$s, the detection rate should then be satisfyingly fitted by a decreasing exponential with a time constant of 110 $\mu$s.

\begin{figure}[ht!]
       \centering    
       \includegraphics[width=0.5\textwidth]{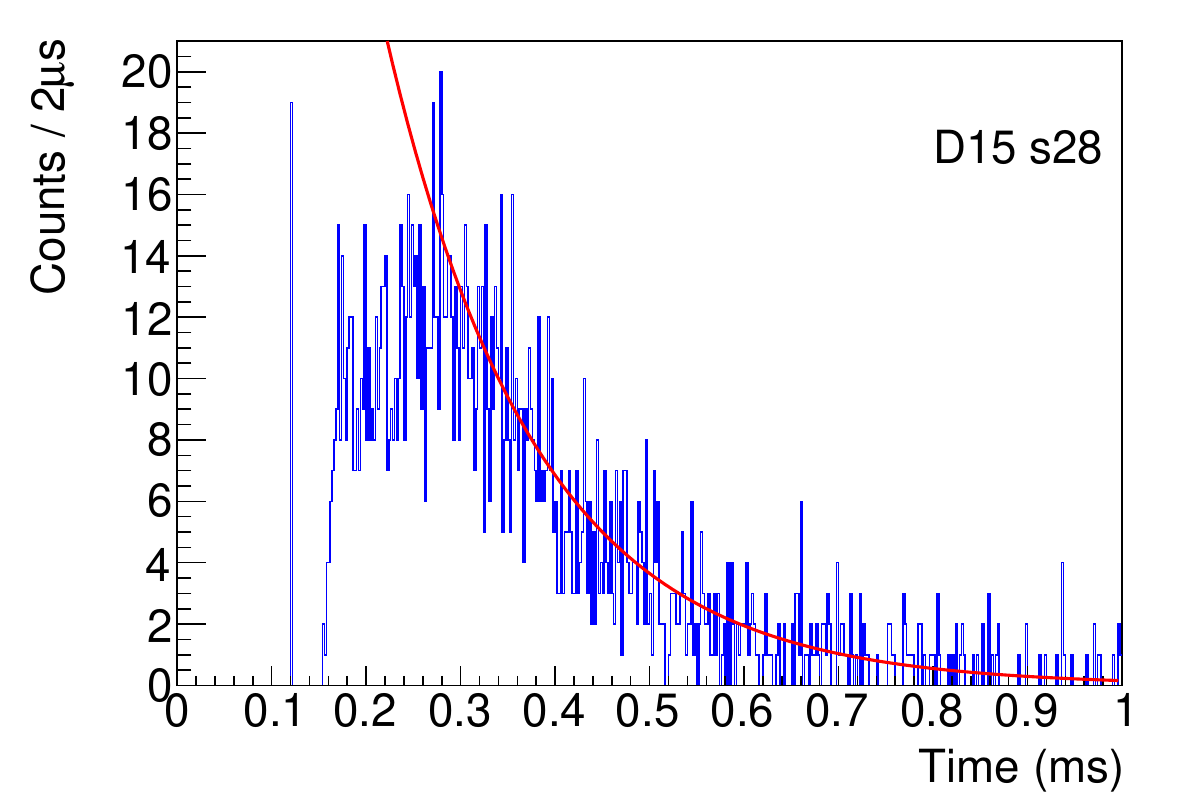}
       \caption{Histogram of the light collected by the detector unit 15 according to the time during shot 28. The red curve represents a decreasing exponential with a half-life of 110 $\mu$s.}
       \label{fig:time_s28}
\end{figure}

As shown in Fig.\ref{fig:time_s28}, we see that, after the prompt gamma flash peak and some dead time, the peak detection rate increases until reaching a maximum and later decreases. If we represent a decreasing exponential with a 110 $\mu$s time constant and the right amplitude, we indeed see a satisfactory agreement between this exponential and the detection rate after some time. Using this information, we defined a time cutoff value after which we considered that the detector unit functions satisfyingly.

\begin{figure}[ht!]
      \centering
       \begin{subfigure}[b]{0.49\textwidth}
            \centering
            \includegraphics[width=\textwidth]{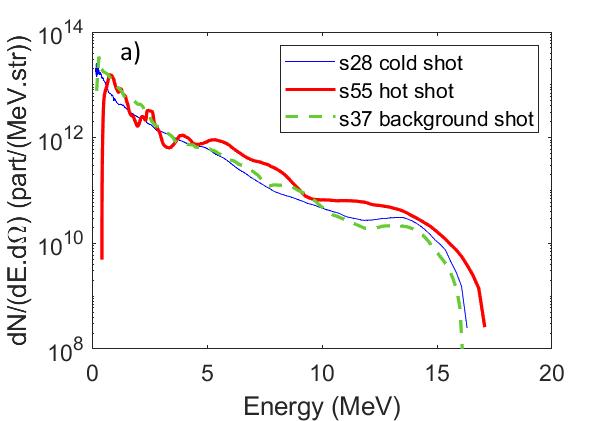}
       \end{subfigure}
       \begin{subfigure}[b]{0.49\textwidth}
            \centering
            \includegraphics[width=\textwidth]{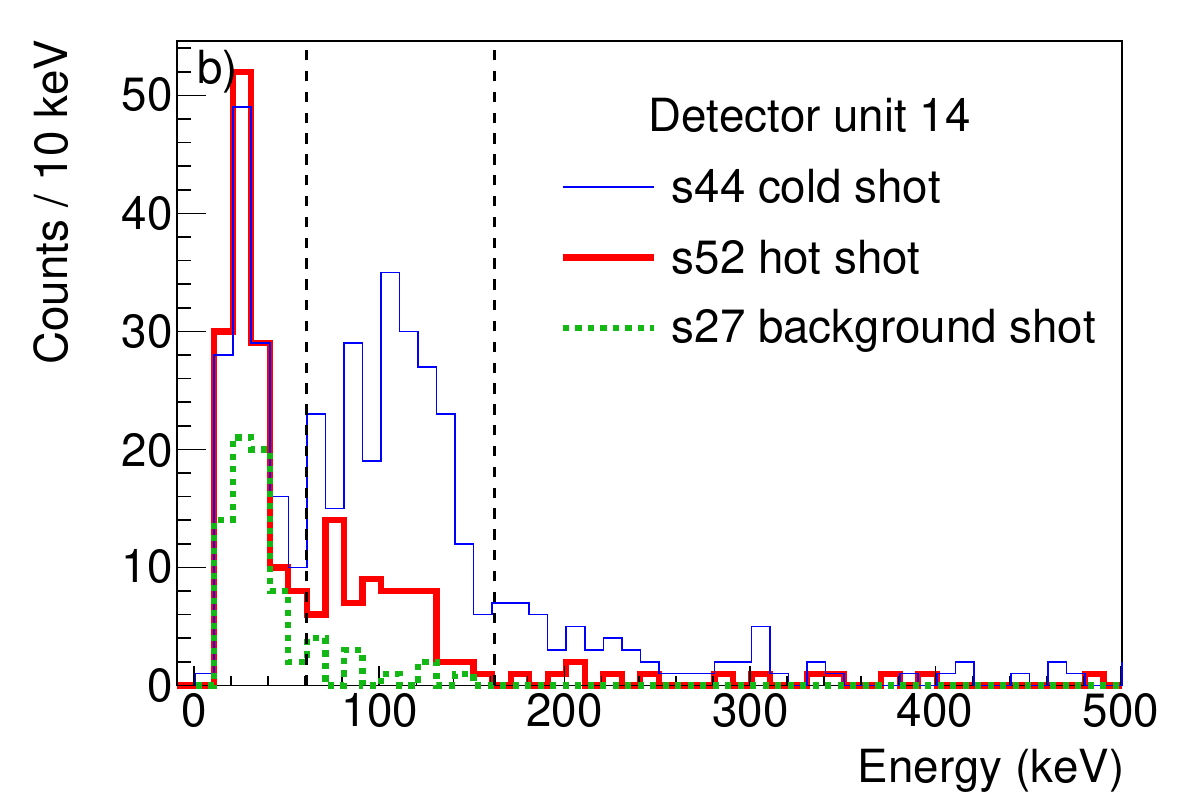}
       \end{subfigure}
       \caption{a) Proton spectra for three shots with equivalent laser parameters (the South beam focal spot is $\sim$150 $\mu$m), one with a solid vanadium secondary target, one without and one with a vanadium secondary target heated by the 100 J ns-duration laser beam. b) Histograms of the light collected by the detector unit 15 according to the energy during those three shots. All peaks detected before 340 $\mu$s are excluded from those histograms.}
       \label{fig:hot_d15}
\end{figure}

Since the background level is different for each PMT and each shot, we also had to set new more adapted thresholds. To do so, we first directly looked at the traces and choose as threshold the half of the mean height of the noise. After that, we tried to improve those thresholds by looking at the histogram where the number of peaks was represented according to its energy, and have the neutron peak as clear as possible above the background. The result of this exercise can be seen in Fig.\ref{fig:hot_d15}b for the different laser configurations.

\subsubsection{Input of Geant4 simulations}

\begin{figure}[ht!]
      \centering
       \begin{subfigure}[b]{0.49\textwidth}
            \centering
            \includegraphics[width=\textwidth]{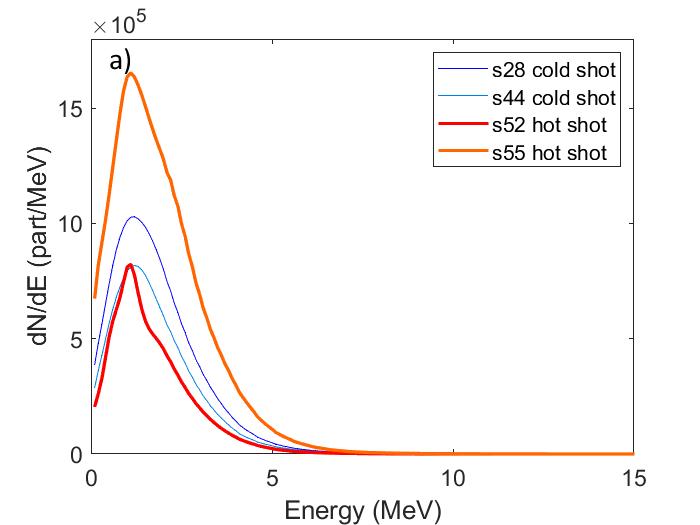}
       \end{subfigure}
       \begin{subfigure}[b]{0.49\textwidth}
            \centering
            \includegraphics[width=\textwidth]{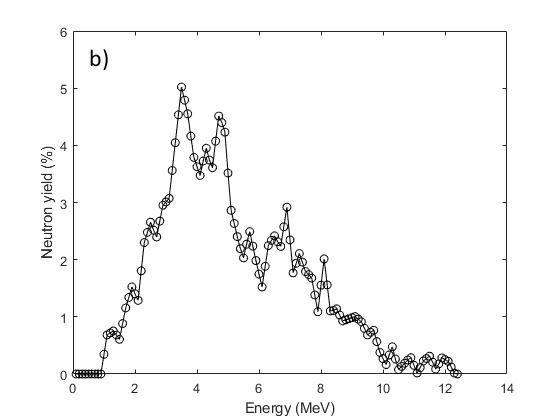}
       \end{subfigure}
       \caption{a) Neutron spectrum calculated from the measured proton spectrum using the TENDL data\cite{tendl}, for the two hot shots 52 and 55 and the two cold shots 28 and 44. b) Neutron spectrum of the PuBe neutron source used for the calibration according to Söderström et al. \cite{soderstrom2021characterization}.}
       \label{fig:neutron_spectrum_shots}
\end{figure}

We used Geant4 \cite{Geant4} simulations to calculate the information needed in order to retrieve an actual number of emitted neutrons from the number of detected neutrons, namely (i) the total efficiency of the neutron detector for a given neutron spectrum, and (ii) the temporal evolution of the detection rate. The neutron energy spectra during the shots were calculated using the measured proton spectra and TALYS package \cite{tendl}, and are shown in Fig.\ref{fig:neutron_spectrum_shots}a. The detector setup has been reproduced in the simulations. A 10 million events simulation was done to retrieve a simulated efficiency for each neutron spectrum. These efficiencies were compared to the efficiency of the detector calculated with Geant4 using the spectrum of the PuBe neutron source used for the calibration (shown in Fig.\ref{fig:neutron_spectrum_shots}b) \cite{lelasseux2021design}, and to the actual measured efficiency with the source. The PET shield and lead shield that can be seen in the middle of the detector array were removed for the simulations using the PuBe neutron source since that calibration stage was done without the presence of those shields.


In order to use accurate cross-sections for scattering and nuclear reactions for neutrons below 20 MeV, we used the NeutronHP physics list. To handle more generic interactions, we also used the physics lists G4DecayPhysics, G4RadioactiveDecayPhysics and G4EmStandardPhysics. Material descriptions are presented in Ref.\cite{lelasseux2021design}. We only considered the deposited energy in the scintillator and not the production of light since attempts to do so using Geant4 have been shown to lead to underestimations \cite{CompositionEJ254,amare2019boronated}. Hence, we considered the Q-value peak, i.e. the energy depositions between 2310 and 2320 keV, as representative of the efficiency. We also considered the neutron sources to be isotropic since the difference between the maximum and minimum calculated angular neutron fluence was only around 3\%.

\begin{figure}[ht!]
       \centering    
       \includegraphics[width=0.4\textwidth]{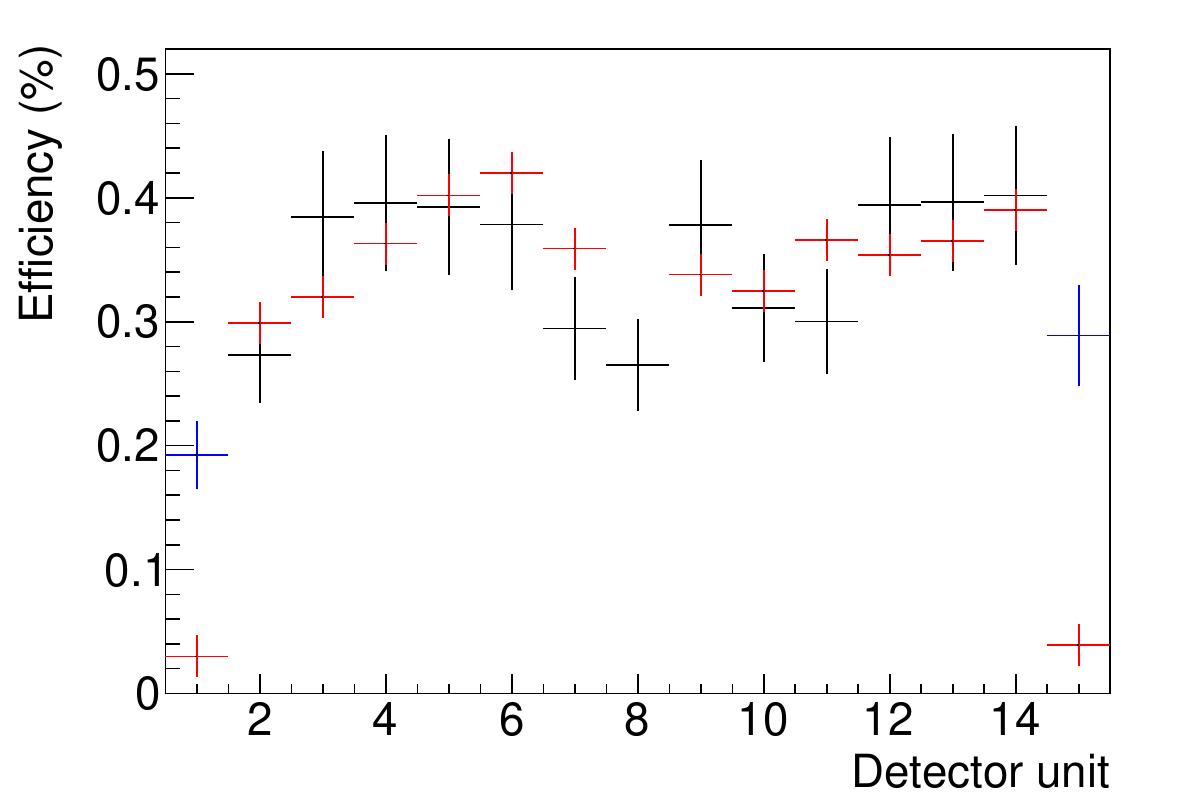}
       \caption{Efficiencies of each detector unit for the PuBe neutron source. The measured efficiencies obtained during calibration are presented in red. The efficiencies calculated thanks to the Geant4 simulations are presented in black for detector units 2 to 14 and in blue for detector units 1 and 15.}
       \label{fig:compare_eff}
\end{figure}

During the calibration (which was performed after the experiment), we noticed that, contrary to what happened during the experiment, the detector units 1 and 15 presented a significant noise level which prevented most of the neutron detections. Hence, we used simulations with a neutron source corresponding to the PuBe neutron spectrum to retrieve a credible efficiency during the experiment for those two units. The detector unit efficiencies are calculated using the following equations :
\begin{align}
    Eff_{unit}=&Eff_{sum}\frac{h_{unit}}{h_{sum}} \\
    Err_{unit}=&Err_{sum}*\frac{h_{unit}}{h_{sum}}+Eff_{sum}*\frac{\sqrt{h_{unit}}}{h_{sum}}\\
    \nonumber &+Eff_{sum}*\frac{h_{unit}}{h_{sum}^2}*\sqrt{h_{sum}}
\end{align}
Where Eff$_{unit}$ is the calculated efficiency for a given unit for the PuBe neutron source, Eff$_{sum}$ is the sum of measured efficiencies for all detector units except 1, 8 and 15, h$_{unit}$ and h$_{sum}$ the 2310-2320 keV peak height obtained during simulations with a given number of events respectively for one detector unit and for all detector units except 1, 8 and 15, and Err$_{unit}$ and Err$_{sum}$ are the errors associated respectively to Eff$_{unit}$ and Eff$_{sum}$. The calculated and measured efficiencies for all detector units for the PuBe neutron source are presented in Fig.\ref{fig:compare_eff}. From now on, the measured efficiencies for detector units 1 and 15 are replaced by the calculated ones.

Using the efficiency related to the PuBe neutron source and the Geant4 simulations presented earlier, we can then calculate efficiencies for each detector unit according to the expected neutron spectrum (showed in Fig.\ref{fig:neutron_spectrum_shots}a). The efficiency values and associated errors have been calculated using the following equations:
\begin{align}\label{eq:eff_simus}
    Eff_{shot}=&Eff_{calib}*\frac{h_{shot}}{h_{calib}} \\
    Err_{shot}=&Err_{calib}*\frac{h_{shot}}{h_{calib}}+Eff_{calib}*\frac{\sqrt{h_{shot}}}{h_{calib}}\\
    \nonumber &+Eff_{calib}*\frac{h_{shot}}{h_{calib}^2}*\sqrt{h_{calib}}
\end{align}

Where Eff$_{shot}$ is the calculated efficiency for a given shot, Eff$_{calib}$ the efficiency related to the PuBe neutron source, h$_{shot}$ and h$_{calib}$ the 2310-2320 keV peak height obtained during simulations with a given number of events and with a neutron emission spectrum corresponding to respectively a shot or the PuBe neutron source, and Err$_{shot}$ and Err$_{calib}$ are the errors associated respectively to Eff$_{shot}$ and Eff$_{calib}$. Those efficiencies are presented in Table.\ref{tab:units_efficiencies}.

\begin{table}[ht!]
    \centering
    \begin{tabular}{|c|c|c|c|c|c|}
        \hline
        Detector unit & PuBe source & Shot 28 & Shot 44 & Shot 52 & Shot 55 \\
        \hline
        1 & 0.192(27) & 0.297(48) & 0.295(48) & 0.303(49) & 0.292(48) \\
        14 & 0.390(44) & 0.372(49) & 0.368(48) & 0.369(48) & 0.358(47) \\
        15 & 0.289(41) & 0.311(50) & 0.315(51) & 0.322(52) & 0.317(51) \\
        \hline
        All & 4.78(62) & 8.49(125) & 8.53(126) & 8.65(128) & 8.36(124) \\
        \hline
    \end{tabular}
    \caption{Calculated efficiencies for every detector unit and shot of interest. Shots 28 and 44 are cold vanadium target shots, while for shots 52 and 55, the vanadium target is ionized.}
    \label{tab:units_efficiencies}
\end{table}

\begin{figure}[ht!]
       \centering    
       \includegraphics[width=0.5\textwidth]{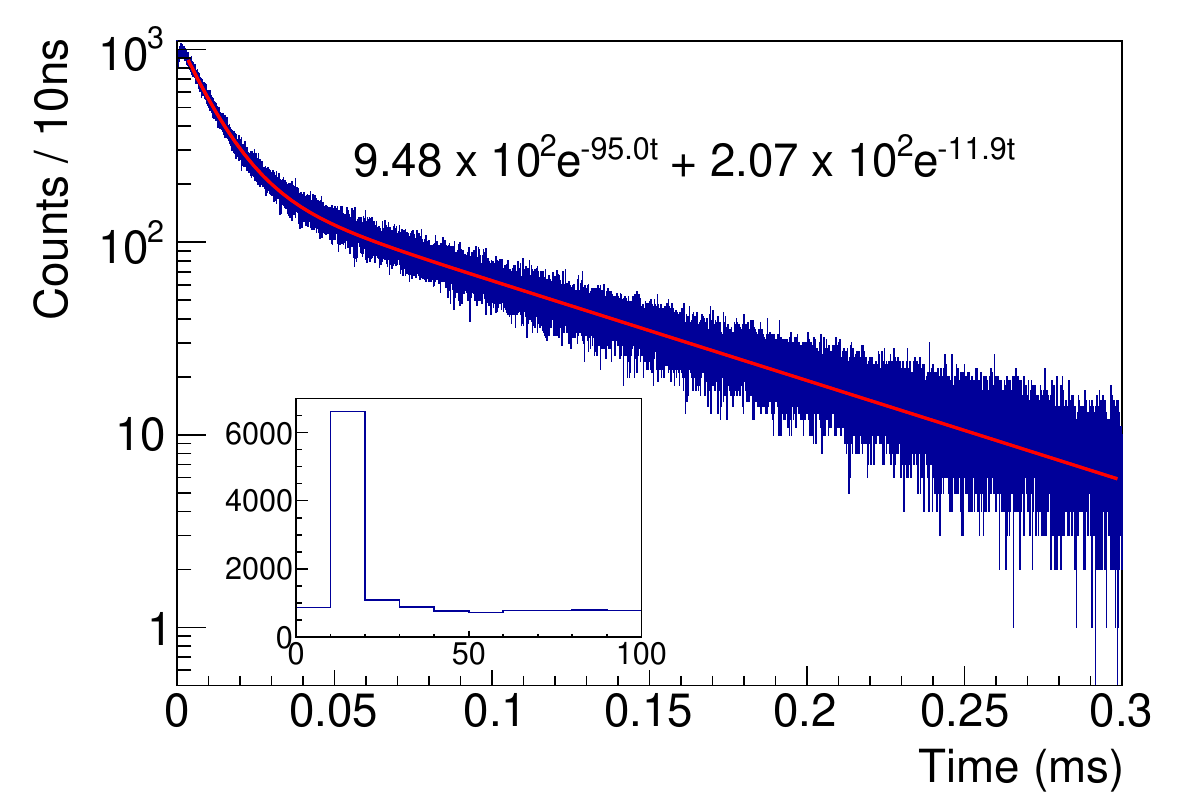}
       \caption{Histogram of energy deposition between 2310 and 2320 keV according to the time of the last interaction depositing energy, from 0 to 300 $\mu$s, according to a Geant4 simulation \cite{Geant4}. The red curve represents the fit and the equation written above corresponds to this fit. The zoom-in view shows the first 100 ns and especially the peak representing mainly events where the neutron capture took place without any rebound in the HDPE.}
       \label{fig:time_simu_s28}
\end{figure}

In order to accurately determine the temporal evolution of the detection rate after hundreds of $\mu$s, we needed to simulate more events. Using the shot 28 spectrum, we simulated the emission of ~180 million neutrons. We decided to use the same temporal evolution for all shots. This temporal evolution and its equation are shown in Fig.\ref{fig:time_simu_s28} for the first 300 $\mu$s.

Using this temporal evolution and the calculated efficiencies, it is possible to retrieve from the count of events in the neutron peaks, like the ones shown in Fig.\ref{fig:hot_d15}b, the number of neutrons produced in the vanadium foil during a shot, using Eqs.\ref{eq:neutron_produced} and \ref{eq:neutron_produced_error}.

\begin{align}
    n_{unit/shot}=&\frac{N_{unit/shot}-N_{unit/background}}{Eff_{unit/shot}}*\frac{f(1)}{f(1)-f(t_{cutoff})} \label{eq:neutron_produced} \\
    errn_{unit/shot} =&\frac{\sqrt{N_{unit/shot}}+errN_{unit/background}}{Eff_{unit/shot}}*\frac{f(1)}{f(1)-f(t_{cutoff})} \nonumber \\ &+\frac{N_{unit/shot}-N_{unit/background}}{(Eff_{unit/shot})^2}*\frac{f(1)}{f(1)-f(t_{cutoff})}*\sqrt{Eff_{unit/shot}} \label{eq:neutron_produced_error}
\end{align}

Where n$_{unit/shot}$ is the number of neutrons produced during the shot as detected by the detector unit, N$_{unit/shot}$ the count of events in the neutron peak for the detector unit during the shot, N$_{unit/background}$ the mean count of events in the neutron peak for the detector unit during background shots and errN$_{unit/background}$ the associated error, Eff$_{unit/shot}$ the efficiency of the detector unit according to the neutron spectrum calculated for the shot, f(t) is the integral from 0 to t of the histogram represented in Fig.\ref{fig:time_simu_s28} with t in ms, t$_{cutoff}$ is the cutoff time used in ms and errn$_{unit/shot}$ is the error associated to n$_{unit/shot}$. As shown in Eq.\ref{eq:neutron_produced_error}, we did not take into account the error induced  by the simulated time evolution of the detection rate. Due to the high statistic in the simulations, this error is negligible compared to the other sources of error.

\section{Results}

Retaining only the six shots for which the proton spectra were similar (see examples in Fig.\ref{fig:hot_d15}a) and the detectors for which every corresponding trace was usable, we find the results presented in Tables \ref{tab:results_same_parameters} and \ref{tab:results_same_parameters2}. During the hot shots 52 and 55, the North beam delivered 100 J on the target.

\begin{table}[]
    \centering
    \begin{tabular}{|c|c|c|c|c|c|c|c|c|c|c|}
        \hline
        Detector & \multicolumn{2}{|c|}{Peak} & Time & \multicolumn{2}{|c|}{Background shots} & \multicolumn{2}{|c|}{Cold shots} & \multicolumn{2}{|c|}{Hot shots} \\
         unit & E$_{min}$ & E$_{max}$ & cutoff ($\mu$s) & s27 & s37 & s28 & s44 & s52 & s55 \\
         \hline
         1 & 80 & 180 & 370 & 17 & 19 & 53 & 72 & 62 & 74 \\
         14 & 60 & 160 & 380 & 13 & 69 & 225 & 229 & 73 & 102 \\
         15 & 80 & 210 & 230  & 54 & 129 & 403 & 427 & 132 & 275 \\
         \hline
    \end{tabular}
    \caption{Neutron detection counts for a given set of analysis parameters for the shots of interest. For the analysis parameters, delimitation of the neutron peak and time cutoffs are presented in this table.}
    \label{tab:results_same_parameters}
\end{table}

\begin{table}[]
    \centering
    \begin{tabular}{|c|c|c|c|c|c|c|c|c|c|}
        \hline
         Detector & \multicolumn{2}{|c|}{Cold shots} & \multicolumn{2}{|c|}{Hot shots} \\
         unit & s28 & s44 & s52 & s55 \\
         \hline
         1 & 6.2(30)x10$^5$ & 9.6(31)x10$^5$ & 7.6(46)x10$^5$ & 1.00(53)x10$^6$ \\
         14 & 2.81(72)x10$^6$ & 2.90(79)x10$^6$ & 4.9(38)x10$^5$ & 9.7(48)x10$^5$ \\
         15 & 1.57(42)x10$^6$ & 1.67(45)x10$^6$ & 2.0(16)x10$^5$ & 9.1(30)x10$^5$ \\
         Estimation & 2.56x10$^6$ & 1.96x10$^6$ & 1.62x10$^6$ & 4.19x10$^6$ \\
         \hline
    \end{tabular}
    \caption{Total number of emitted neutrons according to the measurement of each detector unit for each shot when using conservative parameters. The last row presents estimations based on the measured proton spectra and experimental cross-sections extrapolations from experimental measurements\cite{amare2019boronated}.}
    \label{tab:results_same_parameters2}
\end{table}

It is important to compare the measured number of neutrons with the estimations we can calculate from the measured proton spectrum. Cross-sections from TENDL \cite{tendl} tend to be overestimated for low energies, hence, we decided to use cross sections extrapolated from experimental measurements \cite{amare2019boronated} to get a precise number of produced neutrons. 

\begin{figure}[ht!]
       \centering    
       \includegraphics[width=0.5\textwidth]{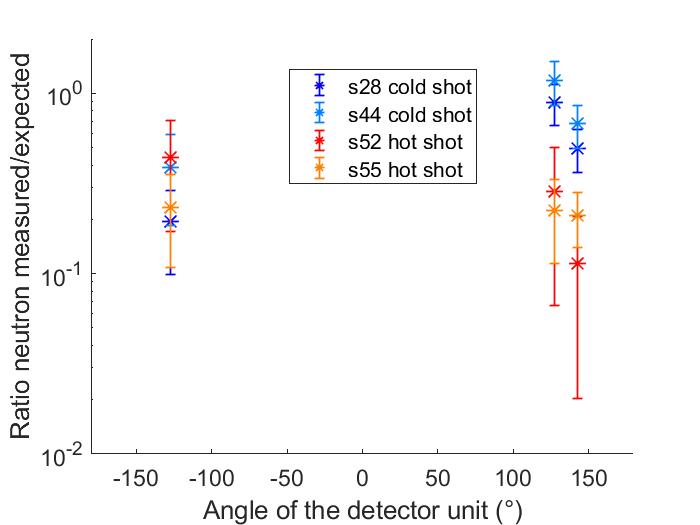}
       \caption{Ratio of the total number of emitted neutrons inferred from the measurements, over the estimated neutron production according to the measured proton spectrum, vs the angle of the detector unit from the picosecond South laser direction.}
       \label{fig:results_ratio_count}
\end{figure}

Using such cross-sections, we calculate the number of neutrons that should have been produced in the secondary vanadium target. Those estimations are shown in the last row of Table.\ref{tab:results_same_parameters2} and the ratio of the measurements over the neutron production expectations is plotted in Fig.\ref{fig:results_ratio_count}. Note that the fact that we used calculated efficiencies for detector unit 1 and 15 does not affect the final result, which is the ratio of neutrons emitted from solid versus plasma targets. As can be seen in Table.\ref{tab:results_same_parameters2} and Fig.\ref{fig:results_ratio_count}, there is a good match between the measurements and expectations for the cold shots. This would tend to validate our methodology. We observed however a significant drop of produced neutrons from the ionized vanadium target, that is compared to what could be expected using the same (p,n) reaction cross-section.

\section{Discussion}\label{sec12}

We will now discuss various potential reasons that could lead to the observed low number of neutrons in the case where the vanadium target is ionized. A first explanation could be a geometrical effect: as the vanadium target expands in vacuum, and as the proton beam is also expanding conically from its source target\cite{bolton2014instrumentation} (as illustrated in Fig.\ref{fig:target_setup}), there will be a reduced amount of vanadium ions that will be intercepted by the protons, compared to the cold case. We estimated quantitatively this effect by doing hydrodynamic simulations of the target expansion using MULTI, conducted in the conditions of the experiment. Using these, we could calculate how many vanadium ions are seen by the proton beam. This led to a $\sim$9.1\% decrease in ions seen by the proton beam in the hot case compared to the cold case. Hence, this cannot explain our observation.

Secondly, we evaluated how much the ionization of the vanadium target could influence the screening of the (p,n) reaction. For this, we estimated the ionization state of the target. As the signal was too low on our X-ray spectrometer for our shot conditions, we raised the North beam energy to 600 J on specific shots, in order to measure the heated vanadium target emission. For these, we measured, at the laser-solid interface, that the heated plasma had a 10$^{21}$ cm$^{-3}$ electron density and a temperature between 1130 and 1200 eV. According to the NIST Atomic Spectra Database\cite{nist}, 1.2 keV corresponds to an ionization rate of 16 for vanadium, the corresponding ionization energy being 1165.2 eV. These values allowed us to anchor our MULTI simulations.

Then, we used MULTI simulations scaled down to our irradiation conditions (100 J) to determine the actual density and ionization level for the shots we analyzed.


Using this density and ionization level, we were able to calculate the evolution of the mean ionization of the target in time. Notably, approximately 2 ns after the start of irradiation by the North beam, this mean target ionization reaches a maximum in time of 12.54. We will consider this maximum value as our ionization in the following in order to look for the maximum possible effect.


 Now, accounting for this target ionization of 12.54, we evaluated how this would impact our neutron production, according to the standard theory \cite{salpeter1954electron,cvetinovic2014electron}. Following it, it would lead to a change of ~0.01\% of the neutron yield, clearly insufficient to account for the observed reduction in the hot plasma.
 However, several publications \cite{rolfs1995status,cvetinovic2020electron,bystritsky2014experimental} present experimental results which suggest that the screening potential value is strongly underestimated by the standard theory. In this frame, Lipoglavsek and Cvetinovic \cite{lipoglavsek2020electron} proposed an empirical formula to calculate the screening potential value following Eq.\ref{eq:Ue_big}.
\begin{equation}\label{eq:Ue_big}
    U_e=Z^2*U_0
\end{equation}
Where U$_e$ is the screening potential, U$_0$=0.8 keV is the maximum electron screening potential measured for the $^2$H(d,p)$^3$H reaction and Z is the electronic cloud charge of the target nucleus. Using this equation, we found new values for the screening potential, of 423.2 keV for the solid vanadium and 87.5 keV for for an ionization of 12.54. This implies a much greater change in the neutron production. The corresponding new estimations for neutron production are given in the second row of Table.\ref{tab:recap_estimation}.


Taking into account a strong screening effect reduces the difference observed between the hot and cold shots, as can be seen by comparing the first two rows of Table.\ref{tab:recap_estimation}, but it is not enough to fully explain it.

Finally, we also considered the difference in stopping power in a plasma or a solid. Since the foil is only 1 $\mu$m thick, we neglected this effect on the measurement of the proton spectrum until now. Since the proton stopping power in a plasma depends on many parameters, we can consider an extreme case to see if this can explain the difference between the hot and the cold shots. We take into account on one side the stopping power in the solid vanadium according to SRIM \cite{SRIM} to rebuild a more accurate emitted proton spectrum. Indeed, since the proton spectrum measurement is performed after passing through the target, the measured proton spectrum is not exactly the same as the emitted one. On the other hand, we consider, as an extreme case, the stopping power in the vanadium in a plasma state as null. Taking into account the important screening potential, the stopping power difference in this extreme case and the plasma expansion, we found the new estimations in the neutron production shown in the fourth row of Table.\ref{tab:recap_estimation}.

As it could be expected, this does not change the resulting ratios significantly and, again, this cannot explain the measured difference between the hot and the cold shots (compare the third and fourth rows of Table.\ref{tab:recap_estimation}).

\begin{table}[]
    \centering
    \begin{tabular}{|c|c|c|c|c|c|c|c|c|c|c|}
        \hline
         Assumptions & \multicolumn{2}{|c|}{Cold shots} & \multicolumn{2}{|c|}{Hot shots} \\
          & s28 & s44 & s52 & s55 \\
         \hline
         Experimental Cross Section (ECS) \cite{amare2019boronated} & 2.56x10$^6$ & 1.96x10$^6$ & 1.62x10$^6$ & 4.19x10$^6$ \\
         ECS + Electron Screening (ES) \cite{lipoglavsek2020electron} & 3.16x10$^6$ & 2.47x10$^6$ & 2.22x10$^6$ & 4.87x10$^6$ \\
         ECS + ES + Stopping Power (SP) & 3.12x10$^6$ & 2.43x10$^6$ & 2.22x10$^6$ & 4.87x10$^6$ \\
         ECS + ES + SP + Plasma Expansion & 3.12x10$^6$ & 2.43x10$^6$ & 2.02x10$^6$ & 4.43x10$^6$ \\
         \hline
         Experimental data & 0.62-2.81 & 0.96-2.90 & 2.0-7.6 & 9.1-10.0 \\
         (range over detectors 1, 14, 15) & x10$^6$ & x10$^6$ & x10$^5$ & x10$^5$ \\
         \hline
    \end{tabular}
    \caption{Estimations of neutron production for the four shots of interest depending on different sets of assumptions, each set containing the assumptions on the line above and a new one, and summary of the experimental measurements (bottom row).}
    \label{tab:recap_estimation}
\end{table}

\section{Conclusion}\label{sec13}

Using the unique capabilities of laser facilities such as LULI2000, we aimed to investigate the influence of the ionization level in plasmas on a nuclear reaction. To do so, we used a (p,n) reaction and a high efficiency neutron detector.
The measurements done for the shots with non-ionized solid target gave satisfactory results in the expected range, based on current experimental knowledge. However, we observed an unexpectedly large drop in the neutron production when the target was ionized. We have shown that classical electron screening, plasma effect on the stopping power or plasma expansion are not enough to explain this decrease, which is yet to be explained.

\section{Acknowledgments}
We acknowledge discussions with I. Pomerantz (Tel-Aviv University) and B. Qiao (Peking University). The authors acknowledge the expertise of the LULI2000 laser facility staff. This work was supported by funding from the European Research Council (ERC) under the European Unions Horizon 2020 research and innovation program (Grant Agreement No. 787539, Project GENESIS) and by Grant No. ANR-17-CE30-0026-Pinnacle from the Agence Nationale de la Recherche. We acknowledge, in the framework of Project GENESIS, the support provided by Extreme Light Infrastructure Nuclear Physics (ELI-NP) Phase II, a project co-financed by the Romanian Government and the European Union through the European Regional Development Fund, and by the Project No. ELI-RO-2020-23, funded by IFA (Romania), and by the Romanian Ministry of Research and Innovation under research contract PN~23~21~01~06, to design, build, and test the neutron detectors used in this project.

\section{Data Availability}
All data needed to evaluate the conclusions in the paper are present in the paper. Experimental data are archived on servers at LULI and are available from the corresponding author upon reasonable request.






\bibliography{sn-bibliography}


\end{document}